\documentclass{article}
\usepackage{amsmath}
\usepackage{amssymb}

\begin{document}

\begin{center}
\LARGE 
\textbf{A Generalized Quantum Theory}
\\[1 cm]
\normalsize
Gerd Niestegge
\\[0,3 cm]
\scriptsize
Fraunhofer ESK, Hansastr. 32, 80686 M\"unchen, Germany
\\
gerd.niestegge@esk.fraunhofer.de,
gerd.niestegge@web.de
\\[1 cm]
\normalsize
\end{center}
\normalfont \itshape Abstract. \normalfont 
In quantum mechanics, the selfadjoint Hilbert space operators play a triple role
as observables, generators of the dynamical groups and
statistical operators defining the mixed states. 
One might expect that this is typical of Hilbert space quantum mechanics,
but it is not.
The same triple role occurs for the elements of a certain ordered Banach space
in a much more general theory based upon
quantum logics and a conditional probability calculus
(which is a quantum logical model of the L\"uders-von Neumann measurement process).
It is shown how positive groups, automorphism groups, Lie algebras
and statistical operators
emerge from one major postulate - the non-existence of
third-order interference (third-order interference 
and its impossibility in quantum mechanics
were discovered by R. Sorkin in 1994).
This again underlines the power of the combination 
of the conditional probability calculus
with the postulate that there is no third-order interference.
In two earlier papers, its impact on contextuality and nonlocality had already 
been revealed.
\\[0,3 cm]
\textit{Key Words.} Foundations of quantum mechanics, dynamical groups, positive groups, Lie algebras, operator algebras
\\[0,3 cm]
\textit{PACS.} 03.65Fd, 03.65:Ta, 02.30.Tb

\section{Introduction}

In quantum mechanics, the selfadjoint Hilbert space operators play a triple role. 
First of all, they represent the observables which are the physically measurable 
quantities of the system under consideration. Second, the normalized positive 
trace-class operators are called statistical operators; they define the mixed 
states of the system. Third, the selfadjoint operators are generators of 
one-parameter dynamical groups describing reversible time evolutions of the system.

One might expect that this triple role is typical of Hilbert space quantum
mechanics. In the present paper, it will be shown that this is not true. 
Such a triple role occurs in a much more general theory based upon
quantum logics and a conditional probability calculus which is a
quantum logical model of the L\"uders-von Neumann measurement process. 
This theory has been elaborated by the author in some recent papers
\cite{Nie01, Nie08, Nie10, Nie12, Nie13, Nie14}.

A further major assumption is required; this is the absence of third-order interference.
The concept of third-order interference was introduced by Sorkin
who also recognized that third-order interference is ruled out by quantum mechanics \cite{Sor94}.
His concept was adapted to conditional probabilities by Barnum, Emerson and Ududec \cite{Udu11}.

This paper does not consider the role of the observables in the generalized 
quantum theory, since it was already sufficiently studied 
in \cite{Nie08}. The statistical operators are addressed
briefly. The paper focuses on the group generators,
using the theory of order derivations introduced
by Connes \cite{Con74}.

In sections 2 and 3, the conditional probability calculus and Sorkin's
concept of third-order interference are recapped as far as needed in
this paper. 
First new results concerning the statistical operators and trace states
are presented in section 4.
Order derivations are briefly sketched in section 5,
and positive semigroups are considered in section 6, before then
turning to the major new results.
These are the dynamical groups and Lie algebras emerging when
third-order interference is ruled out (sections 7, 8, 9). 
Section 10 is dedicated to equivalent reformulations
of some of the mathematical conditions used so far by means of 
the conditional probabilities; this makes them  
accessible to physical interpretation. 
In the last two sections,
it is shown how Jordan algebras and von Neumann algebras fit in
the generalized theory. 

\section{The conditional probability calculus}

In quantum mechanics, the measurable quantities of a physical 
system are re\-presented by observables. Most simple are those 
observables where only the two discrete values 0 and 1 are 
possible as measurement outcome; these observables are called \textit{events} (or \textit{propositions})
and are elements of a mathematical structure called \textit{quantum logic}.

A quantum logic $E$ contains two specific elements $0$ and $\mathbb{I}$ 
and possesses an \textit{orthogonality relation} $\bot$,
an \textit{orthocomplementation} $E\ni e \rightarrow e' \in E$
and a \textit{partial sum operation} + which is defined 
only for orthogonal events.
Moreover, $e' \bot e$ and $e + e' = \mathbb{I}$ for $e \in E$.
The interpretation of this mathematical terminology is as follows: 
orthogonal events are exclusive, $e'$ is the negation of $e$, and
$e + f$ is the conjunction or and-function of the two 
exclusive events $e$ and $f$.

The states on a quantum logic are the analogue of the 
probability measures in classical probability theory, and 
conditional probabilities can be defined similar to 
their classical prototype \cite{Nie01, Nie08, Nie10}. 
A \textit{state} $\mu$ allocates the probability $ \mu(f) \in [0,1]$ to each 
event $f$, 
is additive for orthogonal events, and $\mu(\mathbb{I})=1$.
The \textit{conditional 
probability} of an event $f$ under another event $e$ is the 
updated probability for $f$ after the outcome of
a first measurement has been the event $e$; it is denoted 
by $ \mu(f | e) $. Mathematically, it is defined by the
conditions that the map $E \ni f \rightarrow \mu(f | e)$
is a state on $E$ and that the identity 
$ \mu(f | e) = \mu(f)/\mu(e)$ holds for all events 
$f \in E$ with $f \bot e'$. It must be assumed that $\mu(e) \neq 0$.

However, among the abstractly defined quantum logics, 
there are many where no states or no conditional 
probabilities exist, or where the conditional probabilities 
are ambiguous. Therefore, only those quantum logics where 
sufficiently many states and unique conditional 
probabilities exist can be considered a satisfying 
framework for a probabilistic theory.
\newpage

In \cite{Nie08, Nie12}, it has been shown that 
such a quantum logic $E$ generates an order-unit space $A$
(partially ordered real linear space with a specific norm; see \cite{AS01})
and can be embedded in its unit interval $\left[0,\mathbb{I}\right]$
$:=$ $\left\{a \in A : 0 \leq a \leq \mathbb{I} \right\}$ 
$=$ $\left\{a \in A : 0 \leq a\ and\ \left\| a \right\| \leq 1 \right\}$;
$\mathbb{I}$ becomes the order-unit, and
$e' = \mathbb{I} - e$ for $e \in E$. 
Each state $\mu$ on $E$ has a unique positive linear extension on $A$ 
which is again denoted by $\mu$.

Let $K$ denote the state space consisting
of all states of the quantum logic $E$, and let $V$ be the real-linear space
generated by $K$. A norm can be defined on $V$ such that $V$ becomes a 
base-norm space. The order-unit space $A$ is the dual of $V$,
and the unit ball of $A$ is compact with regard to the weak
topology $w(A,V)$. $A$ is the weakly closed linear hull of $E$.

Note below that an operator $S: A \rightarrow A$ on the order-unit space $A$ is called positive
if $S(a) \geq 0$ for all $a$ in $A$ with $a \geq 0$. Most interesting are 
the positive operators $S$ with $S(\mathbb{I}) = \mathbb{I}$; 
the reason is that, in this case, the map
$E \ni e \rightarrow \mu(S(e))$
defines a state $\mu S$ on $E$ for any state $\mu$ on $E$, and
the map $S^{*}: \mu \rightarrow \mu S$ becomes a transformation of the state space $K$. 

As shown in \cite{Nie08, Nie12}, for each event $e$ in $E$, there is a weakly continuous positive linear operator 
$U_e : A \rightarrow A$ with the following properties:
$\mu(f|e)\ \mu(e) = \mu(U_e f)$ for all $f \in E$ and all states $ \mu $,
$\mu(U_e x) = \mu(x)$ for all $x \in A$ and any state $\mu$ with $\mu(e)=1$,
$\mu(U_e x) = 0$ for all $x \in A$ and any state $\mu$ with $\mu(e)=0$,
$U_e^{2} = U_e$, $e = U_e e = U_e \mathbb{I}$, $0 = U_e f$ as well as $U_e U_f = 0$ 
for $f \in E$ with $e \bot f$, and
$f = U_e f$ for $e' \bot f$. 

These positive projections $U_e$ have many similarities with the compressions
considered by Alfsen and Shultz \cite{AS03} and called P-projections in their earlier papers.
However, the two concepts differ; $U_e x = 0$ with an event $e$ and a positive element $x$ in $A$
does not imply $U_{e'} x = x$ and, therefore, $U_e$ is not a compression (P-projection).
Moreover, Alfsen and Shultz's major interest are the spectral convex sets, but
the results of this paper will show that a rich theory might also be possible without
assuming spectrality.

Further weakly continuous linear operators $T_e$ and $S_e$ can now be defined for each $e \in E$ by 
$T_e(x) := \frac{1}{2} (x + U_e x - U_{e'}x)$ and $S_e(x) := 2U_e(x) + 2 U_{e'}x - x$, $x \in A$.
The properties of the operators $U_e$ above imply the following properties for these operators: 
$\mu(T_e x) = \mu(x)$ for all $x \in A$ and any state $\mu$ with $\mu(e)=1$, and
$\mu(T_e x) = 0$ for all $x \in A$ and any state $\mu$ with $\mu(e)=0$. Moreover,
$e = T_e e = T_e \mathbb{I}$, $0 = T_e f$ for $f \in E$ with $e \bot f$, $S_e^{2}x = x$ 
for any $x$ in $A$, $T_e + S_e T_e = 2 U_e $ and $U_e = 2T_{e}^{2} - T_e$ ($e \in E$). 

In the remaining part of this paper, it shall always be assumed that $E$ is a
quantum logic with the conditional probability calculus as described in this section.
An interesting link between the linear operators $T_e$ and Sorkin's concept 
of third-order interference shall be considered in the following section.

\section{Third-order interference}

Sorkin \cite{Sor94} introduced the following mathematical term $I_3$ 
for a triple of pairwise orthogonal events $ e_1 $, $ e_2 $ and $ e_3 $, 
a further event $f$ and a state $ \mu $:
$$
\begin{array}{rcl}
  I_3 & := & \mu(f | e_1 + e_2 +e_3) \, \mu(e_1 + e_2 + e_3) - \mu(f | e_1 + e_2) \, \mu(e_1 + e_2) \\
    &   &    \\
    &   & - \mu(f | e_1 + e_3) \, \mu(e_1 + e_3) - \mu(f | e_2 + e_3) \, \mu(e_2 + e_3) \\
    &   &    \\   
    &   & +  \mu(f | e_1) \, \mu(e_1) + \mu(f | e_2) \, \mu(e_2) + \mu(f | e_3) \, \mu(e_3) \\
\end{array}
$$
He recognized that $ I_3 = 0 $ is universally valid in 
quantum mechanics. His original de\-fi\-ni\-tion refers to probability measures 
on `sets of histories'. Using conditional pro\-ba\-bi\-li\-ties,
$ I_3 $ gets the above shape, which was seen by Ududec, Barnum and 
Emerson \cite{Udu11}. 

For the three-slit set-up considered by Sorkin, the identity  $I_3 = 0$ 
means that the interference pattern observed with three open slits is a simple 
combination of the patterns observed in the six different cases when only one 
or two of the three slits are open. 
The new type of interference which is present whenever $ I_3 \neq 0 $ 
holds is called \textit{third-order interference}. 

In Ref. \cite{Nie12}, it has been shown that the quantum logic $E$
rules out third-order interference ($I_3 = 0$)
if and only if the identity
$T_{e+f} x = T_e x + T_f x$
holds for all orthogonal event pairs $e$ and $f$ in $E$ and all $x$ in $A$.
Mathematically, this orthogonal additivity of $T_e$ in $e$ is a lot easier to handle than the 
equivalent identity $I_3 = 0$ with the above definition of the rather intricate 
term $I_3$ which, however, may be more meaningful physically. 

Quantum logics which do not exhibit third-order interference
(i.e., which satisfy the identity  $ I_3 = 0 $) have been studied 
in Ref. \cite{Nie12}, and it has been shown that
there is a product operation  $\Box$ in the order-unit 
space $A$ generated by such a quantum logic, if the 
$\epsilon$-Hahn-Jordan decomposition property holds in addition (\cite{Nie12} Lemma 10.2). 

The quantum logic $E$ is said to possess the $\epsilon$-Hahn-Jordan decomposition property
if, for every bounded orthogonally additive real-valued function $\rho$ on $E$ and every $\epsilon > 0$,
there are two states $\mu$ and $\nu$, nonnegative real numbers $s$ and $t$ and an event $e$ in $E$ such that
$\rho = s \mu - t \nu$ and $\mu(e) < \epsilon$ as well as $\nu(e') < \epsilon$.
It implies that $\left[0,\mathbb{I}\right]$ is the
weakly closed convex hull of $E$ \cite{Nie12}.

The product $ a \Box b  $ is linear and weakly continuous in $a$ as well as in $b$ and 
satisfies the inequality 
$\left\| a \Box b \right\|  \leq  \left\| a \right\| \: \left\| b \right\| $ ($a,b \in A$),
where $ \left\| \:  \right\| $ denotes the order-unit norm on $A$.
For any events $e$ and $f$ in $E$, the identity $T_e f = e \Box f$ holds. 
The events $e$ become idempotent elements in $A$ (i.e., $e = e^{2} = e \Box e$), 
and $e \Box f = 0 $ for any orthogonal event pair $e$ and $f$.
Generally, however, the product is neither commutative nor associative.  
Moreover, the square $a^{2} = a \Box a$ of an element $a$ in $A$ need not be positive.

\section{Statistical operators}

An element $\mu \in K$ is called a \textit{trace state} if 
$\mu (f) = \mu(f | e) \mu (e) + \mu(f | e') \mu (e')$
holds for all events $e,f$ in $E$. This means that all events
are compatible under $\mu$ \cite{Nie10}. 
With the linear operators $U_e$, $T_e$ and $S_e$
defined in section 2, it follows that $\mu$ is invariant under
$U_e + U_{e'}$ and $S_e$ for each event $e$ in $E$. The identity
$T_e + S_e T_e = 2U_e$ then gives $ \mu (T_e f) = \mu (U_e f) \geq 0$
for any events $e,f$ in $E$.

Suppose now that the $\epsilon$-Hahn-Jordan decomposition property
and $I_3 = 0$ hold. For the then existing product and a trace state $\mu$, we get
$\mu(e \Box f) = \mu (T_e f) \geq 0$ for any events $e$ and $f$,
and therefore $\mu(x \Box y) \geq 0$ for any positive 
elements $x$ and $y$ in $A$. 
Note that the $\epsilon$-Hahn-Jordan decomposition property
implies that $\left[0,\mathbb{I}\right]$ is the
weakly closed convex hull of $E$ \cite{Nie12}.

With a trace state $\mu$, each positive element
$x$ in $A$ with $\mu (x) = 1$ then gives rise to two further states: 
$e \rightarrow \mu(x \Box e)$ and $e \rightarrow \mu(e \Box x)$. They 
become identical if the product is commutative. Via this construction, the 
positive elements of the order unit space $A$ define states in the same way
as the statistical operators do in Hilbert space quantum mechanics.
An important question then becomes whether a trace state exists; this shall
now be addressed.

Assume that the linear operator $S_e$ is positive 
for every event $e$ in $E$. Then its inverse $S_e^{-1} = S_e$ is
positive, $S_e(\mathbb{I})=\mathbb{I}$, and these operators generate a 
positive group leaving $\mathbb{I}$ invariant. 
It shall now be
seen that the assumed positivity of the $S_e$ has an important
consequence: the existence of a trace state - at least in the 
finite-dimensional case. The following lemma from \cite{Chu78} will 
be used. 
\\[0,3 cm]
Lemma 1. Let $C$ be a compact convex set in a finite dimensional 
real-linear space. Then the group of all affine homeomorphisms
of $C$ onto $C$ has a common fix point.
\\[0,3 cm]
\textbf{Theorem 1.} Suppose that $S_e$ is positive for each event $e$ in $E$ and
that the dimension of the order unit space $A$ is finite. Then a trace
state exists on $E$.
\\[0,3 cm]
\textit{Proof}. Recall that $K$ is the state space of $E$,
that $V$ is the linear space generated by $K$ and that $A$ is the dual space of $V$.
Since $A$ has a finite dimension, so does $V$ and Lemma 1 can be applied to the 
compact convex set $K$. Therefore, 
there is a common fix point $\mu$ of the affine homeomorphisms
of $K$. 

Since the operators $S_e$ are positive and satisfy $S_e(\mathbb{I})=\mathbb{I}$, 
the transformations $S_e^*$ (defined in section 2) map states to states 
and thus define affine homeomorphisms of $K$.
Therefore, the fix point $\mu$ is invariant under each $S_e$: $\mu(S_e x) = \mu(x)$ 
for any $e \in E$ and $x \in A$. Reconsidering the definition of $S_e$ in section 2, 
this means that $\mu(U_e f) + \mu(U_{e'} f)= \mu(f)$ for any events $e$ and $f$,
and thus $\mu$ is a trace state.
\\[0,3 cm]
Note that Theorem 1 does not require the assumptions that $I_3 = 0$
and the $\epsilon$-Hahn-Jordan decomposition property hold.

\section{Group generators}

A bounded linear operator $D:A \rightarrow A$ is called 
an \textit{order dissipation}, if $e^{tD}$ is positive 
for any $t\geq0$, and is called an \textit{order derivation}, 
if $e^{tD}$ is positive for any real number $t$. The order
dissipations are generators of positive semigroups; each
group element has an inverse which is a linear operator, but
need not be positive. 
The order derivations are generators of positive groups; in 
this case, each group element has a positive inverse.

As described in section 2, the positive operators 
which map $\mathbb{I}$ to $\mathbb{I}$ give rise to transformations of the state space.
Therefore, most interesting are those positive groups, 
which leave the order-unit invariant for all $t$; this holds 
when the generator $D$ satisfies the condition $D(\mathbb{I})=0$.
Such a derivation $D$ generates a one-parameter group
of automorphisms. 
It describes the dynamical evolution satisfying the simple
linear differential equation $ \frac{d}{dt}x_t = Dx_t$ $(x_t \in A)$.
Any physical theory with a reversible time evolution
should include such one-parameter automorphism groups and therefore at least
some derivations $D$ with $D(\mathbb{I})=0$.
Generally, they need not be bounded, but note that only bounded derivations
are considered in this paper. 

The following two lemmas provide useful characterizations of the order dissipations and 
order derivations; the first one is a result in \cite{EHO79} and implies the second one 
which can be found also in \cite{AS01}.
\\[0,3 cm]
Lemma 2. Let $D:A \rightarrow A$ be a bounded linear operator. 
Then the following are equivalent:

(i) $D$ is an order dissipation.

(ii) If $0 \leq x \in A$, $ \mu \in K$ and $\mu (x) = 0$, then $\mu (Dx) \geq 0$. 
\\[0,3 cm]
Lemma 3. Let $D:A \rightarrow A$ be a bounded linear operator. 
Then the following are equivalent:

(i) $D$ is an order derivation.

(ii) If $0 \leq x \in A$, $ \mu \in K$ and $\mu (x) = 0$, then $\mu (Dx) = 0$. 

\section{Positive semigroups}

Now let $P:A \rightarrow A$ be a positive linear operator. Then $P^{n}$ 
is positive for $n = 1,2,3,...$ and, with $P^{0} = I$ 
and $I(x):=x$ for $x$ in $A$,
$$ e^{-t}\ \Sigma ^{\infty}_{n=0} \  \frac{t^{n}}{n!} P^{n} = e^{-t} e^{tP} = e^{t(P-I)}$$
is a convex combination of $I$ and $P^{n}$ ($n = 1,2,3,...$) for any $t\geq0$ and 
therefore positive. This means that $P-I$ is 
an order dissipation. This also follows from Lemma 2.

Therefore, with any $e\in E$, $D:=U_{e} + U_{e'} - I$ is an
order dissipation with $D(1)=0$. Since $U_{e} + U_{e'}$ is idempotent, 
$e^{tD}$ is a simple convex combination of $I$ and $U_{e} + U_{e'}$ and
this case is rather trivial. More interesting is 
$D:=(U_{e} + U_{e'})(U_{f} + U_{f'}) - I$
with a pair of events $e$ and $f$. If 
the linear operators $U_e$ and $U_{e'}$
commute with $U_f$ and $U_{f'}$, the product
$(U_{e} + U_{e'})(U_{f} + U_{f'})$ is an idempotent operator again, yielding
the same trivial situation as above. However, if they don't commute,
this results in a non-trivial positive semigroup 
which leaves the order-unit invariant. Note that, in the classical 
case, $U_e + U_{e'} = I = U_f + U_{f'}$, $D=0$, $e^{tD}=I$ for any $t$,
and the above construction becomes meaningless. 

In the following two sections, it will be seen how
the more interesting positive groups, 
automorphism groups and their Lie algebras
emerge from the absence of third-order
interference in the non-classical case.

\section{Positive groups}

In this section, it is assumed that $E$ rules out third-order interference ($I_3 = 0$)
and satisfies the $\epsilon$-Hahn-Jordan decomposition property.
Then there is a product operation $ \Box $ in $A$ which is neither 
associative nor commutative in the general case.
It will be seen now that, in many cases, the right-hand side multiplication 
operators $ R_{a} : A \rightarrow A $, $R_{a}(x) := x \Box a$ 
are order derivations generating positive groups ($a\in A$). 

One further assumption is required. 
Note that an element in a convex set is an extreme point of this set
if it is not any convex combination of two other elements in this set. 
Denote by $ext[0,\mathbb{I}]$ 
the set of extreme points of $[0,\mathbb{I}]$.
For $\mu \in K$ and $e \in E$ with $\mu(e) = 0$, 
we have from section 2 that $\mu(e \Box x) = \mu (T_e x) = 0$ 
for all $x \in A$.
It shall now be assumed that this holds not only for the
$e \in E$ with $\mu(e) = 0$, but for the $e \in ext[0,\mathbb{I}]$ with $\mu(e) = 0$.
\\[0,3 cm]
\textbf{Theorem 2.} Assume that $\mu(e \Box x) = 0$ for all $x \in A$, if 
$e \in ext[0,\mathbb{I}]$ and $\mu \in K$ with $\mu(e) = 0$.
Then $ R_{a}$ is an order derivation for any $a$ in $A$.
\\[0,3 cm]
\textit{Proof}. Suppose $a \in A$ and $ \mu \in K$. Using Lemma 3, 
it is sufficient to show that  
$$ \left\{ x \in \left[ 0,\mathbb{I} \right] : \mu (x)=0 \right\} \subseteq \left\{ x \in \left[ 0,\mathbb{I} \right] : \mu (R_a x)=0 \right\}. $$
Both sets are convex and weakly compact and, by the Krein-Milman theorem, 
they are the closed convex hulls of their extreme points.
Therefore, it is sufficient to show that any extreme point
of the first set lies in the second one. 

For any extreme point $e$ of the first set, suppose that 
$e = s b_1 + (1-s) b_2$ with $0 < s < 1$ and $b_1, b_2 \in \left[0,\mathbb{I}\right]$. Then
$\mu(e)=0$ implies $\mu(b_1)=\mu(b_2)=0$ and both $b_1$ and $b_2$ 
lie in the first set. Since $e$ is an extreme point of this set, 
it follows that $e = b_1 = b_2$. Therefore, $e$ is an extreme point of the unit interval,
thus $\mu (R_a e) = \mu (e \Box a) = 0$,
which means that $e$ lies in the second set.

\section{Automorphism groups}

Most interesting are 
the order derivations $D$ with $D(\mathbb{I}) = 0$, since then
the positive groups they generate leave the order-unit $\mathbb{I}$ invariant 
and give rise to transformation groups of the state space.
This case shall now be studied. 
Assume again that $E$ rules out third-order interference ($I_3 = 0$), that $E$
satisfies the $\epsilon$-Hahn-Jordan decomposition property and that 
$\mu(e \Box x) = 0$ for all $x \in A$, if 
$e \in ext[0,\mathbb{I}]$ and $\mu \in K$ with $\mu(e) = 0$.

An order derivation $D$ is called \textit{skew}, if $D(\mathbb{I}) = 0$, 
and is called \textit{selfadjoint}, if there is an element $a \in A$ with $D = R_a$.
Of course, $R_a(\mathbb{I}) = a$.  
This naming (selfadjoint and skew) is rather unmotivated here,
but will become clear later when the von Neumann algebras 
will be considered as an example.

Any order derivation $D$ is the sum of a
selfadjoint order derivation $D_1$ and a skew order derivation $D_2$;
with $a := D(\mathbb{I})$ choose $D_1 := R_a$ and $D_2 := D - D_1$.
The commutator $D_0 := [D_1,D_2] = D_1 D_2 - D_2 D_1$
of any two order derivations $D_1$ and $D_2$ is an 
order derivation again and the order derivations 
form a Lie algebra \cite{AS01}. 

It is obvious that 
the commutator is skew if $D_1$ and $D_2$ are skew.
Therefore the skew order derivations form a Lie subalgebra $L$
which shall be called the Lie algebra of the quantum logic $E$. Its elements are 
generators of one-parameter automorphism groups 
which describe reversible dynamical evolutions.
With any pair of elements
$a$ and $b$ in the order-unit space $A$, the operator $\left[R_a,R_b\right] - R_d$
with $ d := b \Box a - a \Box b $ now lies in the Lie algebra $L$ by Theorem 2.
The associativity of the product $\Box$ would imply that $\left[R_a,R_b\right] - R_d = 0$;
however, it is not associative generally.

\section{The commutative case}

The question whether not only the right-hand side multiplication
operators $R_a$, but also the 
left-hand side multiplication operators $T_a: A \rightarrow A$,
$T_a x := a \Box x$ are order derivations for $a \in A$,
shall now be addressed; it will turn out that they are so 
if and only if the product $\Box$ is commutative.
Note that these $T_e$ with $e \in E$ coincide 
with the linear operators $T_e$ considered in section 2.

The following lemma holds under the general assumptions of section 2 and does not
require the further assumptions concerning third-order interference, 
the Hahn-Jordan decomposition property and the extreme points of $\left[0,\mathbb{I}\right]$.
\\[0,3 cm]
Lemma 4. If the operators $T_e - T_{e'}$ and $T_f - T_{f'}$ 
are order derivations for two events $e,f \in E$,
then the identity $T_e f = T_f e$ holds.
\\[0,3 cm]
\textit{Proof}. 
Assume that $D_e$ := $T_e - T_{e'}$ = $U_e - U_{e'}$ 
and $D_f$ := $T_f - T_{f'}$ = $U_f - U_{f'}$
are order derivations for the two events $e,f \in E$, and
define the positive operators
$P_e := U_e + U_{e'}$ and $P_f := U_f + U_{f'}$. Note that $D_e^2 = P_e$ 
and $D_f^2 = P_f$.

Then $\mu (U_{f'} U_f x) = 0$ for any $\mu \in K$ and $0 \leq x \in A$,
since $U_{f'} U_f = 0$. Applying Lemma 3 to the derivation
$D_e$, the positive linear functional $\mu U_{f'}$ and the positive
element $U_f x$ in $A$, it follows that $\mu (U_{f'} D_e U_f x) = 0$ 
for any $\mu \in K$ and $0 \leq x \in A$. Therefore $U_{f'} D_e U_f = 0$.
Similarly $U_f D_e U_{f'} = 0$. 
An immediate consequence is
$$ (U_f - U_{f'}) (U_e - U_{e'}) (U_f - U_{f'}) = (U_f + U_{f'}) (U_e - U_{e'}) (U_f + U_{f'})$$
(both sides being equal to $U_f D_e Uf + U_{f'} D_e U_{f'}$). 
Clearly the same equality holds with exchanged roles of $e$ and $f$. Thus
\begin{center}
$ D_f D_e D_f = P_f D_e P_f, \ D_e D_f D_e = P_e D_f P_e$ and
\end{center}
\hspace*{0,5 cm} $(D_e D_f - D_f D_e)^{2}\mathbb{I}$ 
\\[0,3 cm]
\hspace*{2 cm} $= D_e D_f D_e D_f \mathbb{I} + D_f D_e D_f D_e \mathbb{I} - D_e {D_f}^{2} D_e \mathbb{I} - D_f {D_e}^{2} D_f \mathbb{I}$
\\[0,3 cm]
\hspace*{2 cm} $= D_e D_f D_e D_f \mathbb{I} + D_f D_e D_f D_e \mathbb{I} - D_e P_f D_e \mathbb{I} - D_f P_e D_f \mathbb{I}$
\\[0,3 cm]
\hspace*{2 cm} $= D_e D_f D_e D_f \mathbb{I} + D_f D_e D_f D_e \mathbb{I} - D_e P_f D_e P_f \mathbb{I} - D_f P_e D_f P_e \mathbb{I}$
\\[0,3 cm]
\hspace*{2 cm} $= 0.$
\\[0,3 cm]
In the second but last line, the identities $P_e \mathbb{I} = \mathbb{I} = P_f \mathbb{I}$ have been used to replace $\mathbb{I}$ by 
$P_e \mathbb{I}$ and $P_f \mathbb{I}$, respectively, and the last line follows from the identity above. Therefore
$$e^{t \left( D_e D_f - D_f D_e \right)} \mathbb{I} = \mathbb{I} + t \left( D_e D_f - D_f D_e \right) \mathbb{I}.$$
Since the set of order derivations is closed under commutators \cite{AS01}, 
$D_e D_f - D_f D_e$ is an order derivation and the left-hand side of the last equation
is positive for all $t$. This implies 
\\[0,3 cm]
\hspace*{1,75 cm} $0 = \left( D_e D_f - D_f D_e \right) \mathbb{I} $
\\[0,3 cm]
\hspace*{2 cm} $= (T_e - T_{e'})(f - f') - (T_f - T_{f'})(e - e')$
\\[0,3 cm]
\hspace*{2 cm} $= (T_e - T_{e'})(2f - \mathbb{I}) - (T_f - T_{f'})(2e - \mathbb{I})$
\\[0,3 cm]
\hspace*{2 cm} $= 2 T_e f - 2 T_{e'}f - e + e' - 2 T_f e + 2 T_{f'} e + f - f'$
\\[0,3 cm]
\hspace*{2 cm} $= 2 T_e f - 2f + 2 T_{e}f - e + e' - 2 T_f e + 2e - 2 T_{f} e + f - f'$
\\[0,3 cm]
\hspace*{2 cm} $= 4 T_e f - 4 T_f e $.
\\[0,3 cm]
In the second but last line, the identities $T_e + T_{e'} = I = T_f + T_{f'}$ have been used.
Therefore $T_e f = T_f e$.
\\[0,3 cm]
\textbf{Theorem 3.} Suppose that $I_3 = 0$ and 
the $\epsilon$-Hahn-Jordan decomposition property hold and that
$\mu(e \Box x) = 0$ for all $x \in A$, if 
$e \in ext[0,\mathbb{I}]$ and $\mu \in K$ with $\mu(e) = 0$.
Then the following are equivalent:

(i) For any $a \in A$, the operator $T_a$ is an order derivation.

(ii) The product $\Box$ is commutative.
\\[0,3 cm]
\textit{Proof}. Assume (i). Then particularly the operators $T_{e - e'} = T_e - T_{e'}$ 
are order derivations for the events $e$ in $E$, and Lemma 4 implies $T_e f = T_f e$
for any two events $e$ and $f$ in $E$. This means $e \Box f = f \Box e$. Since the product
is linear and weakly continuous in each component and $A$ is the weakly closed linear hull of $E$,
the product is commutative.

Now assume (ii).
This means that $T_a = R_a$ for $a \in A$,
and the $T_a$ become order derivations by Theorem 2.
\\[0,3 cm]
If the product $\Box$ is commutative, the commutator 
$\left[R_a,R_b\right] = \left[T_a,T_b\right]$ is  a skew 
order derivation for any two elements $a$ and $b$ in $A$
and thus lies in the Lie algebra $L$. In this case, $L = \left\{0\right\}$
would imply that the operators $T_a$, $a \in A$, 
and particularly the $T_e$, $e \in e$,
commute with each other. Then the $U_e$ would commute
and it would follow that, with any $e,f \in E$, 
$U_e f + U_{e'} f = U_e U_f \mathbb{I} + U_{e'} U_f \mathbb{I} =
U_f U_e \mathbb{I} + U_f U_{e'} \mathbb{I} = U_f e + U_f e' = U_f \mathbb{I} = f$.  
This would mean that all events in the quantum logic $E$
would be compatible \cite{Nie10} and $E$ would be classical. 
Vice versa, as soon as there are two events
which are not compatible, the Lie algebra $L$ is not trivial.

\section{Some equivalent reformulations by means of the conditional probabilities}

The property that the operators $T_e - T_{e'}$ 
are order derivations for all events $e$ has 
been studied by Iochum and Shultz under the name
\textit{ellipticity} in a more specific setting  
in order to characterize the state spaces of the JBW algebras
among the spectral convex sets \cite{AS03, IoS83}.
The proof of Lemma 4 is a simple transfer of the proof of 
Theorem 9.48 in \cite{AS03} to the more general setting of this paper.

Ellipticity is a mathematical property which has no immediate 
physical or probabilistic interpretation. The next lemma
presents an equivalent property which is more accessible 
to interpretations.
\begin{tabbing}
Lemma 5. Under the general assumptions of section 2, the following
are equi-\\
valent for any event $e$ in the quantum logic $E$:\\
\hspace*{0,5 cm}(i) \hspace*{0,1 cm} \= $T_e - T_{e'}$ is an order derivation.\\
\hspace*{0,5 cm}(ii) \> $\mu(f) - \mu(f|e)\mu(e) - \mu(f|e')\mu(e')$ \= $\geq  -2 \sqrt{\mu(f|e)\mu(e) \mu(f|e')\mu(e')}$ \= for all \\ 
\hspace*{1 cm} \> events $f$ and all states $\mu$.\\
\hspace*{0,5 cm}(iii) \> $ \mu(f) - \mu(f|e)\mu(e) -  \mu(f|e')\mu(e') $ \> $ \leq 2 \sqrt{\mu(f'|e)\mu(e) \mu(f'|e')\mu(e')}$ \> for all \\
\hspace*{1 cm} \> events $f$ and all states $\mu$.
\end{tabbing}
Proof. (i) $\Leftrightarrow$ (ii): Note that $T_e - T_{e'} = U_e - U_{e'}$, $(U_e - U_{e'})^{2} = U_e + U_{e'}$ and $(U_e - U_{e'})( U_e + U_{e'}) = U_e - U_{e'}$.
Therefore, $(U_e - U_{e'})^{n} = U_e - U_{e'}$ for $n = 1,3,5,...$, $(U_e - U_{e'})^{n} = U_e + U_{e'}$ for $n = 2,4,6,...$ and
\\[0,3 cm]
\hspace*{2 cm} $ exp\left(t\left(T_e - T_{e'}\right)\right) = \sum^{\infty}_{n=0} \frac{t^{n}}{n!} (U_e - U_{e'})^{n}$
\\[0,3 cm]
\hspace*{3,55 cm} $= I + \sum^{\infty}_{n=1} \frac{t^{n}}{n!} U_e + \sum^{\infty}_{n=1} \frac{(-t)^{n}}{n!} U_{e'}$
\\[0,3 cm]
\hspace*{3,55 cm} $= I + \sum^{\infty}_{n=0} \frac{t^{n}}{n!} U_e + \sum^{\infty}_{n=0} \frac{(-t)^{n}}{n!} U_{e'} - U_e - U_{e'}$
\\[0,3 cm]
\hspace*{3,55 cm} $= I + exp(t)U_e + exp(-t)U_{e'} - U_e - U_{e'}$
\\[0,3 cm]
for any real number $t$. Positivity of this operator means that 
$$0 \leq \mu \left( f + exp(t)U_e f + exp(-t)U_{e'} f - U_e f - U_{e'} f \right)$$ 
for all events $f$ and states $\mu$. That is
$$0 \leq \mu(f) + exp(t) \mu(f|e)\mu(e) + exp(-t) \mu(f|e')\mu(e') - \mu(f|e)\mu(e) - \mu(f|e')\mu(e').$$
Now note that, with any two nonnegative real numbers $\alpha$ and $\beta$, the largest lower bound for the function 
$t \rightarrow \alpha \: exp(t) + \beta \: exp(-t)$ is $2\sqrt{\alpha \beta}$. Therefore, 
the last inequality holds for all real $t$ if and only if
$$0 \leq \mu(f) + 2 \sqrt{\mu(f|e)\mu(e) \mu(f|e')\mu(e')} - \mu(f|e)\mu(e) - \mu(f|e')\mu(e').$$
(ii) $\Leftrightarrow$ (iii): 
Replacing $f$ by $f'$ in (ii) gives:
$$1 - \mu(f) - \mu(e) + \mu(f|e)\mu(e) - \mu(e') + \mu(f|e')\mu(e')
\geq -2 \sqrt{\mu(f'|e)\mu(e) \mu(f'|e')\mu(e')}$$
and $ \mu(f) - \mu(f|e)\mu(e) -  \mu(f|e')\mu(e') 
\leq  2 \sqrt{\mu(f'|e)\mu(e) \mu(f'|e')\mu(e')}$,
\\[0,3 cm]
which completes the proof.
\\[0,3 cm]
With Lemma 5, ellipticity becomes a feature of the conditional probabilities and imposes 
important restrictions on the typical quantum interference which is exhibited by
the violation of the classical identity $\mu(f) = \mu(f|e)\mu(e) +  \mu(f|e')\mu(e')$,
$e,f \in E$ and $\mu \in K$. Lemma 5 (ii) and (iii)
provide a lower bound and an upper bound
for the interference term $\mu(f) - \mu(f|e)\mu(e) - \mu(f|e')\mu(e')$.
 
The other condition in Lemma 4 can also be equivalently reformulated
by means of the conditional probabilities in the following way.
\\[0,3 cm]
Lemma 6. Under the general assumptions of section 2, the following
are equivalent for any two events $e$ and $f$ in the quantum logic $E$:
\newline
\hspace*{0,5 cm}(i) $T_e f = T_f e$.
\newline
\hspace*{0,5 cm}(ii) $\mu(f'|e)\mu(e) + \mu(f|e')\mu(e') = \mu(e'|f)\mu(f) + \mu(e|f')\mu(f')$ for all states $\mu$.
\\[0,3 cm]
Proof. Condition (ii) means $U_e f' U_{e'}f = U_f e' + U_{f'} e$ and this is equivalent to $T_e f = T_f e$.
\\[0,3 cm]
Under the assumptions of Theorem 3, condition (ii) or (iii) of Lemma 5 and condition (ii) of Lemma 6 become equivalent, 
since condition (i) of Lemma 5 and condition (i) of Lemma 6 are equivalent then. The equivalence
of these properties of the conditional probabilities will be hard to see directly - without considering 
the order-unit space $A$ and the operators $U_e$ and $T_e$ on $A$ ($e \in E$). The same holds 
for the implications in the following corollary.
\\[0,3 cm]
Corollary 1. Under the general assumptions of section 2, the quantum logic $E$
rules out third-order interference, whenever it satisfies one of the 
following four conditions:

(i) $T_e - T_{e'}$ is an order derivation for each $e \in E$ (i.e., the state space $K$ is elliptic).

(ii) $\mu(f) - \mu(f|e)\mu(e) - \mu(f|e')\mu(e') \geq  -2 \sqrt{\mu(f|e)\mu(e) \mu(f|e')\mu(e')}$ 
for all events $e,f \in E$ and all states $\mu \in K$.

(iii) $ \mu(f) - \mu(f|e)\mu(e) -  \mu(f|e')\mu(e') \leq  2 \sqrt{\mu(f'|e)\mu(e) \mu(f'|e')\mu(e')}$ 
for all events $e,f \in E$ and all states $\mu \in K$.

(iv) $\mu(f'|e)\mu(e) + \mu(f|e')\mu(e') = \mu(e'|f)\mu(f) + \mu(e|f')\mu(f')$ 
for all events $e,f \in E$ and all states $\mu \in K$.
\\[0,3 cm]
Proof. By Lemma 4, 5 and 6, each one of the conditions (i), (ii), (iii) and (iv) implies
$T_e f = T_f e$ for any $e,f \in E$. Now suppose $e_1, e_2 \in E$ with $e_1 \bot e_2$.
Then $T_{e_1 + e_2} f = T_f (e_1 + e_2) =$ $T_f e_1 + T_f e_2 = T_{e_1} f + T_{e_2} f$
for all $f \in E$ and thus $T_{e_1 + e_2} x = T_{e_1} x + T_{e_2} x$ for all $x \in A$. 
This is equivalent to $I_3 = 0$ for all states and events (see section 3).
\\[0,3 cm]
The positivity of the operators $S_e$, $e \in E$, which plays an important role in section 4,
shall now be reconsidered. Again
there is an equivalent property of the conditional probabilities.
\\[0,3 cm]
Lemma 7. Under the general assumptions of section 2, the following
are equivalent for any event $e$ in the quantum logic $E$:
\newline
\hspace*{0,5 cm}(i) $S_e$ is positive. 
\newline
\hspace*{0,5 cm}(ii) $\mu(f) \leq 2\mu(f|e)\mu(e) + 2\mu(f|e')\mu(e')$ for all events $f$ and states $\mu$.
\\[0,3 cm]
Proof. Condition (ii) means $I \leq 2(U_e + U_{e'})$. That is $0 \leq S_e$.
\\[0,3 cm]
More information concerning the physical interpretation 
can be found in Refs. \cite{AS01, AS03, Nie12} for condition (ii) of Lemma 6
and in Ref. \cite{Nie12} for condition (ii) of Lemma 7.

\section {Jordan algebras and Lie algebras}

The formally real Jordan algebras were introduced by Jordan, von Neumann and Wigner \cite{JNW34}. 
Much later, this theory was extended to include infinite dimensional algebras;
these are the so-called JB-algebras and JBW-algebras \cite{AS03}.

The idempotent elements of a JBW-algebra $A$ 
form a quantum logic $E$. 
In this case, $E = ext \left[0,\mathbb{I}\right]$ holds.
If the Jordan algebra does not contain a 
direct summand of type $I_2$, $E$ possesses a conditional probability calculus \cite{Nie01}. 
With the so-called triple product 
$\left\{x,y,z\right\} := x \circ (y \circ z) - y \circ (z \circ x) + z \circ (x \circ y)$,
then $U_e a = \left\{ e,a,e\right\}$, $T_e a = e \circ a$, and $S_e a = \left\{e - e',a,e - e' \right\}$
for any $a \in A$ and $e \in E$. The operators $S_e$ are positive.

Moreover, third-order interference is ruled out, and the $\epsilon$-Hahn-Jordan
decomposition property is satisfied. The product $\Box$ coincides with the Jordan product $\circ$ 
and is commutative.
The order automorphisms leaving the unit $\mathbb{I}$ invariant coincide with the Jordan automorphisms.
Each property of the conditional probabilities considered in section 10 is satisfied.

There are three classes of simple
formally real Jordan algebras with finite dimension and one further case. 
These are the hermitian $n$$\times$$n$-matrices with real, complex and quaternion 
entries, equipped with the usual Jordan product, and the 
hermitian $3$$\times$$3$-matrices with octonion entries \cite{AS03}. The Lie algebras
of the quantum logics consisting of the idempotent elements
of these Jordan algebras are $\mathfrak{so}(n)$, $\mathfrak{su}(n)$, $\mathfrak{sp}(n)$ 
and the exceptional Lie algebra $\mathfrak{f}_4$.

However, there are four further exceptional simple Lie algebras with finite dimension
($\mathfrak{g}_2$, $\mathfrak{e}_6$, $\mathfrak{e}_7$ and $\mathfrak{e}_8$) 
to which no formally real Jordan algebra can be
allocated \cite{Baez01}. An important question now becomes whether they are the Lie 
algebras of some unknown quantum logics. In the case of a positive answer, it would
be interesting to study the characteristics which distinguish them from the 
quantum logics emerging from the formally real Jordan algebras.
Considering the results in \cite{Nie08, Nie12},
it is very likely that their state spaces are not spectral 
(in the meaning of Alfsen and Shultz \cite{AS03}). 
 
\section {Von Neumann algebras}

Quantum mechanics uses a very special type of quantum logic $E$; 
it consists of the selfadjoint projection operators on a 
Hilbert space or, more generally, in a von Neumann algebra $M$.
The selfadjoint part of a von Neumann algebra $M$ is a JBW-algebra $A$.
In this case, 
$U_e a = eae$, $T_e a = (ea + ae)/2$, $S_e a = (e - e')a(e - e')$ 
and moreover $e^{tR_b} a =e^{tb/2} a e^{tb/2}$
($a,b \in A$, $e \in E$).
Again each property of the conditional probabilities considered in section 10 is satisfied.

Furthermore, $D_b a := i(ba - ab)/2$ ($a \in A$) defines a 
skew order derivation $D_b$ for any $b$ in $A$ and 
$e^{tD_b} a =e^{itb/2} a e^{-itb/2}$ ($a \in A$, $t \in \mathbb{R}$).
This specific relation between the elements of $A$ and skew order derivations
distinguishes those JBW-algebras that are the selfadjoint part of a
von Neumann algebras from the other JBW-algebras.
Its mathematical formalization is the so-called 
\textit{dynamical correspondence} \cite{AS03}.
Generally, in a JBW algebra $A$, a skew order derivation $D$ can 
be derived via $D := \left[T_a,T_b\right]$ from a pair $a$ and $b$ in $A$
which does not operator-commute, 
but not from a single element in $A$.

\section {Conclusions}

Those quantum logics that entail the conditional probability calculus 
appear to provide a promising generalized quantum theory.
Many of its mathematical properties can be formulated
by means of the conditional probabilities,
which makes them accessible to physical interpretations.

In the present paper,
it has been shown how dynamical groups and Lie algebras emerge 
when third-order interference is ruled out.
This again underlines the power of the combination 
of the conditional probability calculus
with the postulate that there is no third-order interference.
Its impact on contextuality and nonlocality had already 
been revealed in two earlier papers \cite{Nie13,Nie14}. 

An interesting open question now becomes whether 
the four finite-dimension\-al simple exceptional real Lie algebras
that do not arise from the automorphism groups of the formally really Jordan algebras
perhaps arise from the automorphism groups of some unknown quantum logics.

\end{document}